\documentclass[10pt]{article}
\usepackage{amsfonts,amsmath,amssymb}
\usepackage{graphicx,epsf}
\newcommand{\be}{\begin{eqnarray}}
\newcommand{\ee}{\end{eqnarray}}

\renewcommand{\d}{\mbox{{\rm d}}}
\title{\bf Particle Production in Tachyon Condensation}
\author{G.L.~Alberghi\thanks{e-mail: alberghi@bo.infn.it},$\ $
R.~Casadio\thanks{e-mail: casadio@bo.infn.it}
$\ $ and A.~Tronconi\thanks{e-mail: tronconi@bo.infn.it}\\
 \\
{\em Dipartimento di Fisica, Universit\`a di Bologna, and}
\\
{\em Istituto Nazionale di Fisica Nucleare,
Sezione di Bologna, Italy}}
\begin{document}
\maketitle
\begin{abstract}
We study particle production in the tachyon condensation process
as described by different effective actions for the  tachyon.
By making use of invariant operators, we are able to obtain exact results
for the density of produced particles, which is shown to depend strongly
on the specific action.
In particular, the rate of particle production remains finite only for one
of the actions considered, hence confirming results previously appeared
in the literature.
\end{abstract}
%
%
%
\section{Introduction}
\label{intro}
One of the most intriguing features of String Theory is the tachyon
condensation process and its possible description in terms of a tachyon
effective theory.
Finding an effective action for the tachyon field is a difficult task.
Nevertheless, in certain situations (like the decay of non-BPS D-branes)
some aspects of the string dynamics can be described by an effective field
theory action involving only the tachyon and massless modes
(see for example Refs.~\cite{Lambert:2002hk,Lambert:2001fa}).
\par
The form of the effective action for the tachyon depends on the choice of the
region in field space where it can be valid.
One such a region corresponds to the neighbourhood of the perturbative
string vacuum (in which the tachyon $T=0$) where one can reconstruct
the tachyon effective action from string $S$-matrix.
In this case, we expect the effective action to be of the form
\be
\label{Tperturb}
S=-\int \d t\,\d^p x\,
\left( {1\over 2}\,\eta^{\mu\nu}\,\partial_{\mu}T\,\partial_{\nu}T
-{\mu^2\over 2}\,T^2+g\,T^4+\ldots\right)
\ .
\ee
where $\eta_{\mu\nu}$ is the Minkowski metric (Greek indices run from
$0$ to $p$), $\mu$ and $g$ are constants and the dots represent higher
order terms.
\par
A second possibility is trying to reconstruct the effective action near some
exact conformal points.
One such a conformal point is a time-dependent background that represents
an exact boundary conformal field
theory~\cite{Sen:2002an,Sen:2002in,Sen:2002nu,Gutperle:2003xf},
\be
T=f_0\,e^{\mu\,x^0}+{\bar f}_0\,e^{-\mu\,x^0}
\ .
\ee
A special case is then the ``rolling tachyon'' background
\be
\label{rol}
T=f_0\,e^{\mu\,x^0}
\ ,
\ee
where $\mu^2=1$ or $1/2$ in the bosonic and supersymmetric case
respectively, which would be described by the homogeneous
Lagrangian~\cite{Kutasov:2003er}
\be
\label{kutt}
L=-\frac{1}{1+\frac{T^2}{2}}\,
\sqrt{1+\frac{T^2}{2}-\left(\partial_0 T\right)^2}
\ .
\ee
If the above admits a straightforward covariant generalization,
one obtains
\be
\label{tachDBIx}
S=-\int \d t\, \d^p x\, {1\over 1+{T^2 \over 2}}\,
\sqrt{1+{T^2 \over 2}
+\eta^{\mu\nu}\,\partial_{\mu}T\,\partial_{\nu}T}
\ ,
\ee
which has just the form of a ``DBI'' (Dirac-Born-Infeld) action
\be
\label{tachDBI}
S=-\int \d t\,\d^p x\, V(\tilde{T})\,
\sqrt{1+\eta^{\mu\nu}\,\partial_{\mu}\tilde{T}\,
\partial_{\nu}\tilde{T}}
\ ,
\ee
where $\tilde T=\sqrt{2}\,\sinh\left(T/\sqrt{2}\right)$ and
the potential is given by
\be
V(\tilde T)=\left(\cosh\frac{\tilde T}{\sqrt{2}}\right)^{-1}
\ .
\label{Vp}
\ee
Even if the rolling tachyon (\ref{rol}) is an exact solution of the equation
of motion that arises from the action (\ref{tachDBIx}),
it is not completely clear whether such an action should be
used for the description of field theory space around the perturbative
vacuum $T=0$.
In particular, in Refs.~\cite{Berkooz:2002je,Felder:2002sv,Frolov:2002rr},
the stability of the the classical time-dependent solution of
Eq.~(\ref{tachDBIx}) was analyzed, showing that fluctuations become large
in a very short time interval and hence they might significantly change the
classical evolution.
One can also interpret the huge growth of fluctuations as the creation of
a large number of fluctuation modes in the time-dependent background of
the classical solution \cite{Kluson:2003rd}.
We can then expect that at some time near the beginning of the tachyon
condensation the density of the number of particles created reaches the
string density and the linearized approximation, in which fluctuations
are assumed small, breaks down.
\par
Finally, one can also consider as a starting point a tachyon effective
action of the form
\be
\label{Tbos}
S=-\int \d t\,\d^p x\,
\frac{1}{1+T}\sqrt{1+T+
\frac{\eta^{\mu\nu}\,\partial_{\mu}T\,\partial_{\nu}T}{T}}
\ ,
\ee
which arises when one considers the tachyon of a D-brane in bosonic
String Theory and was derived in Ref.~\cite{Kluson2003} following
the proposal of Ref.~\cite{Kutasov:2003er}.
\par
We will analyze the production of particles for the three actions reviewed
above by making use of Lewis' invariant operators~\cite{Lewis}.
This technique yields exact results and will allow us to go beyond
the usual linear approximation in all the cases under consideration.
Our analysis follows the same line as that of Ref.~\cite{Klusonlast}.
In particular, we shall restrict our study of particle production to the
fluctuation modes with initial momentum $k^2>\mu^2$ at infinite past.
Modes with $k^2<\mu^2$ exponentially grow even at the beginning of the
tachyon condensation and this may significantly change the classical
evolution.
The analysis of these modes is a challenging task beyond the scope of
this paper and even for the modes with a positive frequency
$\Omega_k ^2=k^2-\mu^2$ at far past we will get some interesting
results.
\section{Particle Production}
\setcounter{equation}{0}
\label{general}
The fluctuations of the tachyon field obtained by using the effective Lagrangians
in Eqs.~(\ref{Tperturb}), (\ref{tachDBIx}) and (\ref{Tbos}) may be
described by the Hamiltonian of an harmonic oscillator with time dependent
frequency  (as shown in Ref.~\cite{Klusonlast}).
The formalism of Lewis' invariant operators~\cite{Lewis} will then provide us
with the exact expressions for the number of particles produced by the
time-dependence of the frequency.
\subsection{DBI action}
\begin{figure}[t!]
\centering
\epsfxsize=4in
\epsfbox{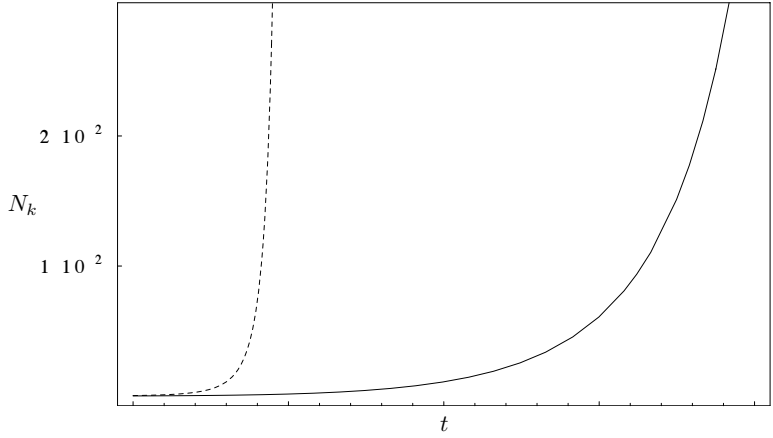}
\caption{{\it Particle density for the linearized approximation
(dashed line) and for the exact solution (solid line)
for the DBI action.}
\label{figura1}}
\end{figure}
Starting with the action given in Eq.~(\ref{tachDBIx}), one can show
that the fluctuations of the tachyon field are governed by the Hamiltonian
for an harmonic oscillator with a time dependent frequency~\cite{Klusonlast}
\be
\Omega^2_{k}(t)={k^2 \over 1+e^{\sqrt{2}\,t}} + m^2
\ ,
\ \ \ \
m^2=- {1 \over 2}
\ .
\label{fre1}
\ee
The expression for the number of created particles is then
\be
\label{particles}
N_k(t)={1 \over {4\,\Omega_k}}\,
\left(\dot \rho^2 + \Omega_k^2\,\rho^2 +{1 \over \rho^2}\right)-{1 \over 2}
\ .
\ee
The auxiliary function $\rho(t)$ is given by
\be
\rho(t)=\sqrt{\frac{x_1^2(t)+x_2^2(t)}{\omega_{0k}}}
\ee
where $\omega_{0k}\equiv\lim_{t\rightarrow-\infty}\Omega_k(t)=k^2-1/2$,
and $x_1(t) $ and $x_2(t) $ are independent solutions of the homogeneous
equation
\be
\ddot x(t) + \Omega_k^2 (t)\,x(t) = 0
\ .
\ee
For the frequency in Eq.~(\ref{fre1}) they are given by
\be
&&
x_1(t) = {1\over 2 }\,\left(y_1 (t) + y_2(t)\right)
\nonumber
\\
\\
&&
x_2(t) = {1\over 2i }\,\left(y_1 (t) -y_2(t)\right)
\ ,
\nonumber
\ee
where
\be
&&
y_1(t)=e^{\strut\displaystyle{i\,\omega_{0k}\,t}}\,
_2F_1\left(-{1 \over 2}+{i \over \sqrt{2}}\,\omega_{0k},
{1\over 2}+{i \over \sqrt{2}}\,\omega_{0k};
1+i\,\sqrt{2}\,\omega_{0k} ; -e^{\sqrt{2}\,t}\right)
\nonumber
\\
\\
&&
y_2(t)=e^{\strut\displaystyle{-i\,\omega_{0k}\,t}}\,
_2F_1\left(-{1 \over 2}-{i \over \sqrt{2}}\,\omega_{0k},
{1\over 2}-{i \over \sqrt{2}}\,\omega_{0k};
1-i\,\sqrt{2}\,\omega_{0k} ; -e^{\sqrt{2}\,t}\right)
\ ,
\nonumber
\ee
and the $F$'s are hypergeometric functions~\cite{abramo}.
One can see from Fig.~\ref{figura1} that the number of produced particles
diverges for a finite $t$ for $ k\gg 1$ as in Ref.~\cite{Klusonlast}, but
this effect seems to be due to the $\Omega_k (t) $
becoming imaginary and not to happen as $ \Omega_k(t)\gg 1$.
In any case, the production obtained by the exact method is significantly
smaller than the result obtained using a linear approximation and seems
to diverge at a later time.
This allows for the possibility that, when backreaction effects are included,
the tachyon condensation process might be correctly described by this
effective action.
\subsection{Bosonic Action}
\begin{figure}[t!]
\centering
\epsfxsize=4in
\epsfbox{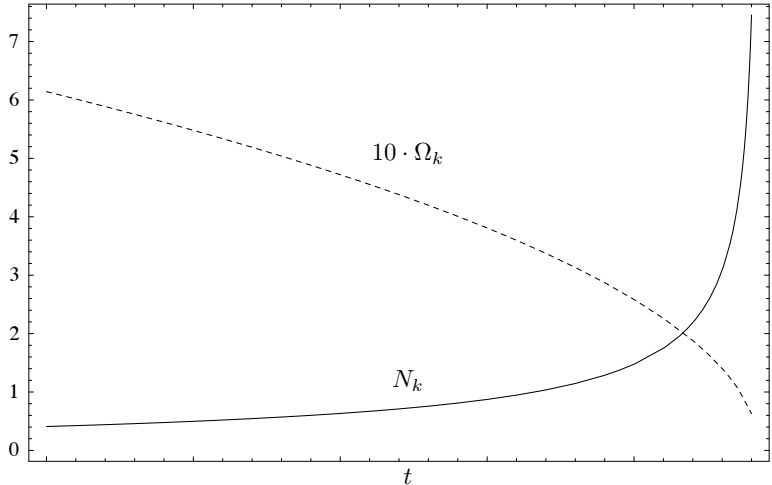}
\caption{{\it Particle density $N_k(t) $ (solid line) and frequency
$ 10\,\Omega_k(t) $ (dashed line) for the Bosonic Action.}
\label{figura2}}
\end{figure}
In this section we will compute the particle production on D-branes in bosonic
String Theory assuming a tachyon effective action of the form given in
Eq.~(\ref{Tbos}).
\par
In the present case the time dependent frequency is given by
\be
\label{omegabos}
\Omega^2 _{k}(t)=\left(1-e^t\right)\,k^2-\frac{1}{4}
\equiv\omega_{0 k}^2-\lambda\,e^t
\ ,
\ee
where $\omega_{0 k}^2=k^2-1/4$ and $\lambda=k^2$.
The expression for the number of produced particles is the same as in
Eq.~(\ref{particles}), where $x(t) $ again solves the homogenous equation
of the harmonic oscillator with frequency $\Omega_k$.
We then have
\be
&&
x_1(t) = {1 \over 2}\,\left[
I_{2\,i\,\omega_{0k}}\left(2\,k\,e^{t/2}\right)\,
\Gamma\left(1 + 2\,i\,\omega_{0k}\right)
+
I_{-2\,i\,\omega_{0k}}\left(2\,k\,e^{t/2}\right)\,
\Gamma\left(1 - 2\,i\,\omega_{0k}\right)
\right]
\nonumber
\\
\\
&&
x_2(t) ={1 \over 2\,i }\,\left[
I_{2\,i\,\omega_{0k}}\left( 2\,k\,e^{t/2}\right)\,
\Gamma\left( 1 + 2\,i\,\omega_{0k}\right)
-
I_{-2\,i\,\omega_{0k}}\left( 2\,k\,e^{t/2} \right)\,
\Gamma\left( 1 - 2\,i\,\omega_{0k} \right)
\right]
\ ,
\nonumber
\ee
where the $I$'s are Bessel functions~\cite{abramo}.
Particle production becomes relevant shortly after the condensation
process has begun, due to the fact that $ \Omega_k (t)$ becomes imaginary,
as is shown in Fig.~\ref{figura2} (in which $\Omega_k$ is multiplied by a
factor of 10 for convenience).
\subsection{Perturbative Effective Action}
\begin{figure}[t!]
\centering
\epsfxsize=4in
\epsfbox{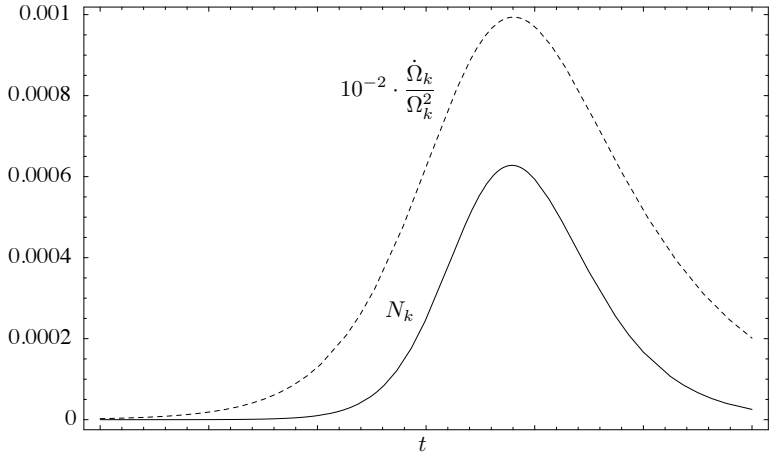}
\caption{{\it Particle density $N_k(t) $ (solid line) and
frequency variation $ \dot \Omega_k(t) / \Omega^2_k (t) $ (dashed line)
for the Bosonic Action.}
\label{figura3}}
\end{figure}
On using the action in Eq.~(\ref{Tperturb}), one obtains the Hamiltonian
of an harmonic oscillator with time dependent frequency
\be
\Omega^2_{k}(t)=
6\,g\,e^{2\,\mu\,t}+k^2-\mu^2\equiv
\lambda\,e^{2\,\mu\,t}+\omega_{0 k}^2
\ ,
\ee
where $\omega_{0 k}^2=k^2-\mu^2$ and $\lambda=6\,g$.
The expression for the number of produced particles is again the same as that in
Eq.~(\ref{particles}), with
\be
&&
x_1(t) = {1 \over 2}\,\left[J_{i\,{\omega_{0k}\over \mu}}\,\left(\sqrt{6\,g\over \mu^2}\,e^{\mu\,t}\right)\,
\Gamma\left(1+ i\,{\omega_{0k}\over \mu}\right)+
J_{-i\,{\omega_{0k}\over \mu}}\left(\sqrt{6\,g\over \mu^2}\,e^{\mu\,t}\right)\,
\Gamma\left(1- i\,{\omega_{0k}\over\mu}\right)
\right]
\nonumber
\\
\\
&&
x_2(t) = {1 \over 2\,i }\,\left[J_{i\,{\omega_{0k}\over\mu}}\left(\sqrt{6\,g\over \mu^2}\,e^{\mu\,t}\right)\,
\Gamma\left(1+ i\,{\omega_{0k}\over\mu}\right)-
J_{-i\,{\omega_{0k}\over\mu}}\left(\sqrt{6\,g \over \mu^2}\,e^{\mu\,t}\right)\,
\Gamma\left(1- i\,{\omega_{0k}\over \mu}\right)
\right]
\ ,
\nonumber
\ee
where the $J$'s are Bessel functions and $\Gamma$ the Gamma function~\cite{abramo}.
\par
We can now restrict the analysys to the bosonic case ($\mu^2=1$),
since for the supersymetric case ($\mu^2=1/2$) the situation is
exactly the same.
The number of produced particles is plotted in Fig.~\ref{figura3}.
One will note that the density of the produced particles reaches a maximum
as the variation of $ \Omega _k (t) $ varies more rapidly and then settles
down.
This behavior strongly suggests that the corresponding action (\ref{Tperturb})
might then be suitable for describing the condensation process up to the time
when the linear approximation breaks down (that is when
$ \mu^2\,T^2 \simeq g\,T^4$).
\section{Conclusions}
\setcounter{equation}{0}
We have computed the density of particles produced in the early stages of
the tachyon condensation.
The use of invariant operators allows us to obtain exact results and
to show that the production rate strongly depends on the form of the
action.
The DBI action~(\ref{tachDBIx}) and the Bosonic Action~(\ref{Tbos})
produce divergent densities in the early stages of the tachyon condensation,
whereas the Perturbative Effective Action~(\ref{Tperturb}) seems to yield a
regular behavior when the linearized approximation is satisfied.
This suggests that such an action should be regarded as the most suitable
candidate to describe the process of tachyon condensation.
On the other hand one should note that backreaction effects have been neglected
in our approach.
Any conclusions therefore need to be confirmed by an analysis that takes
into account the backreaction, and our results must be considered as
preliminary until a more thorough (and very likely numerical) analysis is
performed.
This is left for future research beyond the scope of this paper.
%
%

%
\label{conc}
%

%
%

%
%
%
%

%
\end{document}